\documentclass[letterpaper]{article}
\usepackage{mume}
\usepackage{times}
\usepackage{helvet}
\usepackage{courier}
\usepackage{amsmath}
\usepackage{graphicx}  
\usepackage{subcaption}
\usepackage{hyperref}
\usepackage{amssymb}
\begin{document}
	%
	\title{Query-based Deep Improvisation}
	\author{Shlomo Dubnov \\ University of California in San Diego}
    \maketitle
	\begin{abstract}
	    \begin{quote}
	    In this paper we explore techniques for generating new music using a Variational Autoencoder (VAE) neural network that was trained on a corpus of specific style. Instead of randomly sampling the latent states of the network to produce free improvisation, we generate new music by querying the network with musical input in a style different from the training corpus. This allows us to produce new musical output with  longer-term structure that blends aspects of the query to the style of the network. In order to control the level of this blending we add a noisy channel between the VAE encoder and decoder using bit-allocation algorithm from communication rate-distortion theory. Our experiments provide new insight into relations between the representational and structural information of latent states and the query signal, suggesting their possible use for composition purposes. 
		\end{quote}
	\end{abstract}
	
	\section{Introduction}
    In this paper we introduce the concept of query-based deep musical generation by creating a noisy channel to control the information rate between the input music and the latent features that are derived from the query and are then input into a decoder to produce new music. One of the promises in neural representation of music is the ability to automatically find feature representations that capture essential musical structure without a need for expert human engineering. Moreover, for creative purposes it is important to understand how  structures that were learned for specific musical style can then be used to generate more instances of that musical style in ways that can be controlled or specified by the user. 

	In order to understand how such network operates, we consider a generalization of the Evidence Lower Bound (ELBO) principle of Variational Auto-Encoder (VAE) training in terms of rate-distortion theory that deals with lossy compression and noisy transmission over communication channels. We apply this theory to study aspects of representation learning in VAE, viewed as a source encoder-decoder model. By adding a noisy channel between the encoder-decoder components, the rate of information transfer between the input query and the generator producing music from the noisy latent states can be controlled using a bit-allocation algorithm. Changing the rate of latent information transfer results in different reconstructed musical structures. 

\section{Variational Encoding, Free energy and Rate Distortion}
\label{sec:VA}
	In this work we use the VAE deep learning method to learn a representation of short-term frames in midi files. We have chosen this model for its relative simplicity and generative capabilities. Moreover, the process of learning in VAE is based on so-called free energy minimization principle that makes the complexity-accuracy, or rate-distortion aspect of learning more explicit, as explained below. Controlling the information rate between the input and latent states, or musical surface and features, is normally done during the learning phase by balancing the relative weight of feature complexity penalty
	versus reconstruction accuracy component
	in the loss function \cite{beta-VAE}. 
	
	Rate distortion is a term originating in Shannon's work on lossy communication, measured in terms of mutual information $I(X,Z)$ between a signal $X$ and its compressed version $Z$ under the constraint that the distortion between the two signals $d(X,Z)$ does not exceed a certain threshold. 
	The relation between variational encoding or free-energy approaches and rate distortion were made explicit in the "Broken ELBO" paper \cite{BrokenELBO}. The authors show there that the mutual information between the input $X$ and the latent code $Z$ is bounded below and above by two factors $D$ and $R$ that comprise the ELBO, as follows
	\begin{equation}
	    H-D \leq I_e(X,Z) \leq R
	\end{equation}
	where $H$ is the data entropy, $D$ is the distortion measured as reconstruction log likelihood, and $R$ is the model encoding rate 
	measured by KL divergence between the encoding distribution and the learned marginal approximation. For the sake of completeness, we provide the definitions of the different terms and partial proofs in the appendix. The reader is referred to \cite{BrokenELBO} for more complete mathematical detail. It should be noted that in \cite{BrokenELBO} the authors distinguish between the encoder and decoder distributions, and thus they add subscript 
	\textit{e} to denote specifically the encoding channel. Such distinction unnecessarily complicates our discussion here, so for the most part we will consider them equivalent by Bayes theorem. The important point here is that this expression can be rewritten as a Lagrange optimization by adding $D$ to all terms of the above inequality, giving
	\begin{equation}
	    H \leq I_e(X,Z) + D \leq R + D = -ELBO.
	\end{equation}
	Moreover, allowing for different weight of the distortion $D$, we have
	\begin{equation}
	    I_e(X,Z) + \beta D \leq R + \beta D = -ELBO(\beta).
	\end{equation}
	where $\beta = 1$ gives the original VAE ELBO expression, and $ELBO(\beta)$ corresponds to $\beta$-VAE model.
	This expression is an unconstrained version of the rate distortion,
	\begin{equation}
        \mathcal{L} = I(X,Z) + \beta \langle d(X, Z) \rangle
	\end{equation}
	where minimizing for $\beta$ value gives a point on the rate-distortion curve.
	Changing the value of $\beta$ is important for dealing with two known problems in VAE encoding: Information Preference and Exploding Latent Space problems. The first refers to vanishing of the mutual information between $Z$ and $X$, or in other words $Z$ and $X$ becoming independent due to a powerful decoder. The second problem refers to over-fitting of the output likelihood by matching individually each data point when the training data is finite. 
    In both cases the component of ELBO related to likelihood of the output overshadows the second penalizing factor related to the encoder complexity. 
    
    \subsection{Controlling Information Rate between Encoder and Decoder}
\begin{figure}
        \centering
        \includegraphics[width=0.4\linewidth]
        {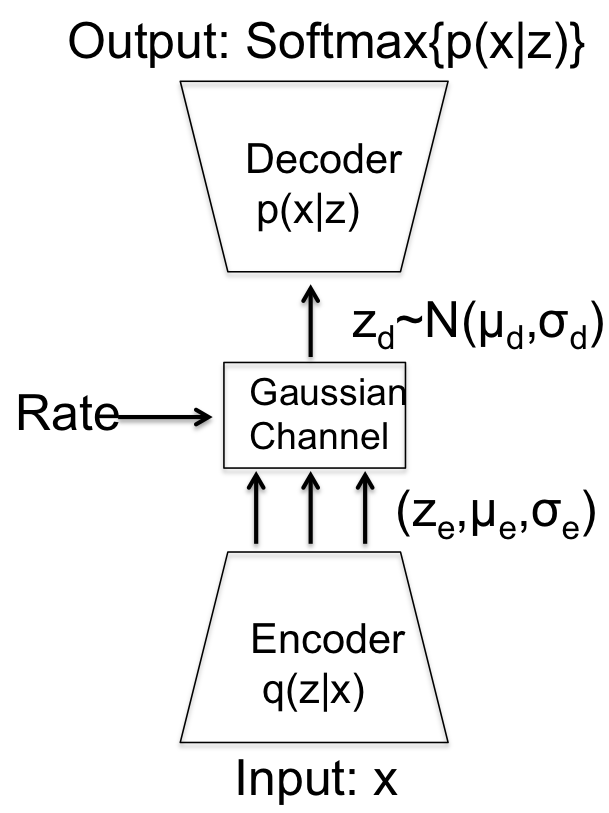}
        \caption{Noisy channel between encoder and decoder}
        \label{fig:AE_noisy_channel}
    \end{figure}

    The above result makes the link between VAE encoding-decoding and rate-distortion theory of noisy communication more evident. During the process of learning, maximization of ELBO effectively reduces the mutual information between the input signal $X$ and the latent state $Z$, while being constrained by quality of reconstruction as measured by $D$. Using this noisy channel model, we consider the connection between musical query and the resulting improvisation through VAE as source encoder and target decoder model. Accordingly, we suggest to add a noisy channel between the source and target, which enables us to introduce information rate control through use of bit-allocation. It is assumed in VAE that the latent states are distributed as multi-variate uncorrelated Gaussians. The rate distortion of such signal is
    \begin{equation}
        R(D) = 
        \begin{cases}
            \frac{1}{2} \log_2 \frac{\sigma^2}{D}, & \text{if $0 \leq D \leq \sigma^2$}\\
            0, & \text{if $D > \sigma^2$}.
        \end{cases}
    \end{equation}
    This rate-distortion can be converted to distortion-rate function $D(R)=\sigma^2 2^{-2R}$, which can be efficiently achieved for a multivariate Gaussian channel by the so called reverse water filling algorithm that starts with a predefined bit-regime and successively allocates one bit at a time to the strongest component, every time reducing the component variance by factor of four and repeating the process until all bits in the bit-pool are exhausted. One should note that channels with variance less then allowed distortion, or channels that run out of bits for a given rate, are given zero bits and thus are eliminated from the transmission.
    We use such channel to reduce the rate of the decoder by adding noise or eliminating some of the weaker latent components. 
    Schematic representation of the channel inclusion in the auto-encoder architecture is given by Figure \ref{fig:AE_noisy_channel}.
    Performing finite bit-size encoding and transmission of the binary quantized latent values from encoder $Z_e$ to decoder $Z_d$ is not required, since we are interested in gating and biasing the original signal towards the prior distribution by encoding it at limited bitrate, which is given by the following optimal channel \cite{Berger}
    \begin{align}
         Q(z_d|z_e) & = Normal(\mu_d,\sigma_d^2) \\
        \mu_d & = z_e + 2^{-2R}(\mu_e - z_e) \\
        \sigma^2_d & = 2^{-4R}(2^{2R}-1)\sigma^2_e
    \end{align}
    
    The practical way of using this channel in our bit-rate controlled model is by plugging each latent encoder value, mean and variance into the above equation, giving us the mean and variance of the decoder's conditional probability, and then picking at random the decoder value according to this distribution. One can see that channels with zero rate will transmit in a deterministic manner the mean value of that element, while channels with infinite rate will transmit the encoder values unaltered. 
    
    
\section{Experimental Results}

    The experiments conducted using our model comprise of querying a pre-trained model with a midi input and reducing the information available to the decoder in a controlled way by passing the encoding through a bit-rate limited channel. Then the decoded music is evaluated both qualitatively for apparent musical structure, and quantitatively in terms of its predictability by analyzing the temporal information rate of the latent states across different time frames. The different experiments that we conducted included training a VAE and then generating new compositions by random sampling of latent states, creating an output sequence by querying the model with another midi file, and finally reconstruction of a midi input query at different channle rates by passing the latent states through a  noisy channel between the encoder and decoder
 
\begin{figure}
        \centering
        \includegraphics[width=\linewidth, height=7cm]
        {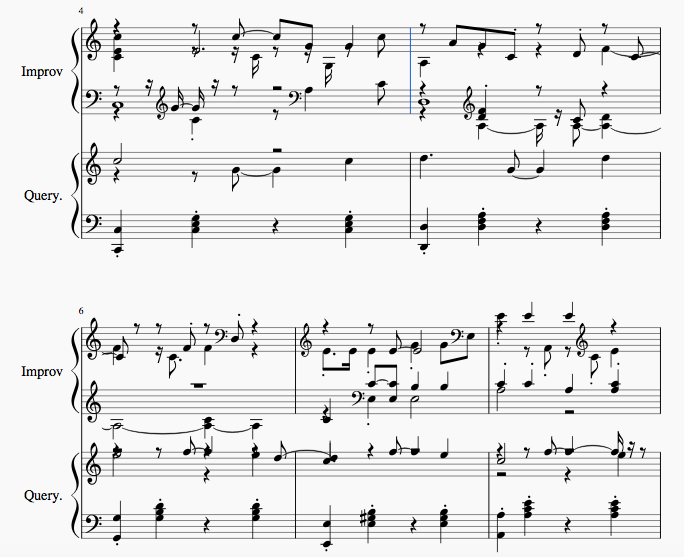}
        \caption{Output of VAE that was trained on Pop Music database, using Naruto Shippuden midi file as an input query. Note that the figure begins with measure four.}
        \label{fig:Naruto-QueryFull}
    \end{figure}

    The VAE was trained on a Pop Music database that contained 126 music clips, mostly comprising of chords and melody, divided into chorus and verse sections. The VAE architecture used here had an input layer comprising of concatention of 16 musical units, each containing notes played at a resolution of 16th notes, representing total of four quarter notes or one bar in 4/4 meter. The hidden layer had 500 units, passed respectively to VAE mean and variance networks for variational encoding.
    As an input query we used midi file from Naruto Shippuden anime. This song is longer and has a different style from the music in the Pop Music corpus that was used for training. Figure \ref{fig:Naruto-QueryFull} shows measures four through eight of the improvised output printed together with the query input. This allows us to analyze their relations.
    We see that the texture of the improvisation is significantly different from the chord-melody texture of the query, but the harmonic relations are preserved and the improvisation reproduces music that matches the chords or the overall harmony of the query.
   
\begin{figure}
        \centering
        \includegraphics[width=\linewidth, height=7cm]
        {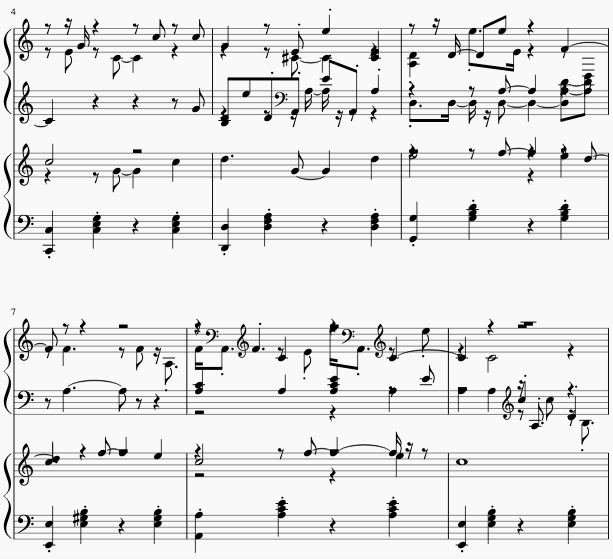}
        \caption{Generation by VAE from a bit-reduced query starting at same measure number four as in the previous figure.}
        \label{fig:Naruto-Query-256}
    \end{figure}

    The correspondence between query and improvisation starts deteriorating when passing through the noise channel. Figure \ref{fig:Naruto-Query-256} shows the results of reducing the bit-rate of the encoder to 256 bits per frame. 
    Musical analysis of the resulting music shows that the improvisation breaks away in some cases from the harmony of the query. For example, in measures four through nine shown in Figure \ref{fig:Naruto-Query-256} , the improvisation shown on top plays a quick chord progression G-A-D, while the query on bottom plays D in the left hand and melody closer to G. Harmonic collisions continue in third and fourth bars, merging together to meet on Amin chord in bar five. 
    Additional bit-reduction results and the uncontrolled random generation are provided online\footnote{\url{https://github.com/sdubnov/qbdi}.}. We can say that further reducing the rate not only makes the improvisation different from query, but also tends to improvise on fewer chords, eventually converging to improvisation on C chord, which is the tonality of the piece. 
    
    \subsection{Analysis of Music Information Dynamics}
    
    Music Information Dynamics broadly quantifies the study of information passing over time between past and future in musical signal \cite{Abdallah}. Study of musical information dynamics was shown to be important for understanding human perception of music in terms of anticipation and predictability. The ability to find repetitions in music depends on the ability to perceive similarity between different variations of musical materials - two similar chords often have different voicing or added notes, melodies modulate, rhythms change, but the overall musical form is still perceivable by capturing unity in this variety. One of the promises in neural modeling of music is the ability to automatically find feature representations that capture essential musical structure without a need for expert human engineering. Moreover, the creative hope is that these structures would be idiomatic to specific style of the training corpus and would allow generating new instances of music in that style in ways that are controllable as desired by the user.
    
    An efficient formal method for studying music information dynamics for signal $X(n)$ is the Information Rate (IR) measure that considers the relation between the present measurement $x=X(n)$ and it's past $\overleftarrow{x}=X(1),X(2),..,X(n-1)$, formally defined in terms of maximum of mutual information between quantized versions of the signal $\hat{X}=Q(X)$
    
    \begin{align}
    IR(X) & = \max_{Q: \hat{X}=Q(X)} I(\hat{x},\overleftarrow{\hat{x}}) \\
    & = H(Q(x))-H(Q(x)|\overleftarrow{Q(x)})
    \end{align}
   
    According to this measure, maximal value of IR is obtained for signals that look uncertain and thus have high entropy $H(Q(x))$, but are predictable or have little uncertainty when the past is taken into account $H(Q(x)|\overleftarrow{Q(x)})$. The need for quantization arises due to the need to detect approximate repetitions in the signal, which in turn depends on the allowed level of signal similarity, or the amount of signal details that is considered when comparing the present to the past. 
    In \cite{VMO} an efficient method called Variable Markov Oracle (VMO) is proposed for estimating IR. This method generalizes a string matching Factor Oracle algorithm to operate over metric space. The quantization is implicitly performed by finding approximate repetitions between time series values up to a threshold. The entropy differences are estimated by considering the compression rate obtained by encoding repeated blocks versus encoding of individual frames. Since a different amount of repetitions is captured at different thresholds, the algorithm finds the highest IR by exhaustive search over all possible thresholds. In addition to finding the numerical IR values, VMO analysis also allows visualization of the salient motifs by enumeration of the salient repetitions and identifying their position and duration in the signal. The motifs are detected by enumerating the repetitions found at optimal thershold and selecting instances that are longer then certain minimal length and have more then minimal number of repetitions. The details of pattern finding are provided in \cite{pattern}
    
    \subsection{Experimental results}
\begin{figure}
        \centering
        \includegraphics[width=\linewidth,  height=5cm]
        {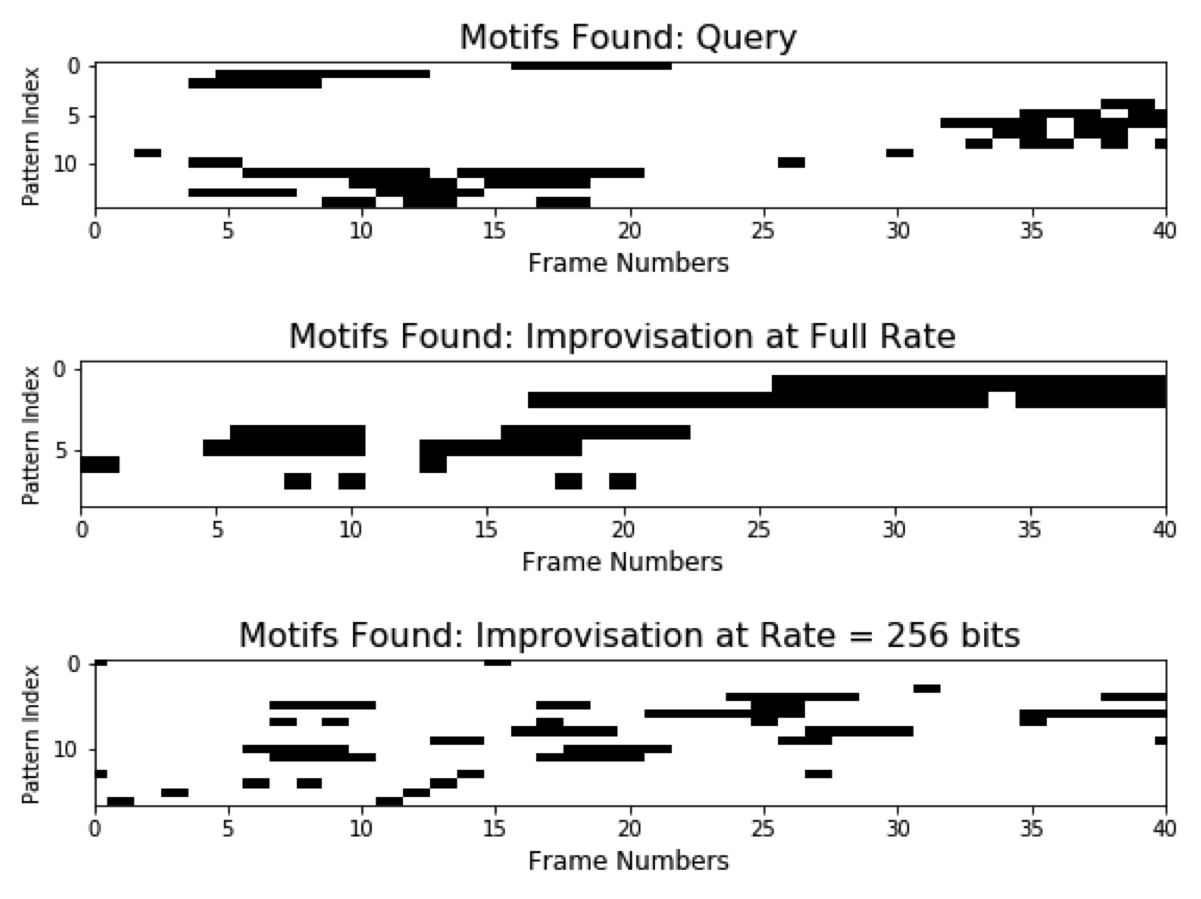}
        \caption{Generation by VAE with bit-rate controlled query: Motifs found in the query (Top), full rate (Middle) and bit-rate limited resynthesis (Bottom). See text for more detail.}
        \label{fig:IR-Motifs}
    \end{figure}

\begin{figure}
        \centering
        \includegraphics[width=\linewidth,  height=5cm]
        {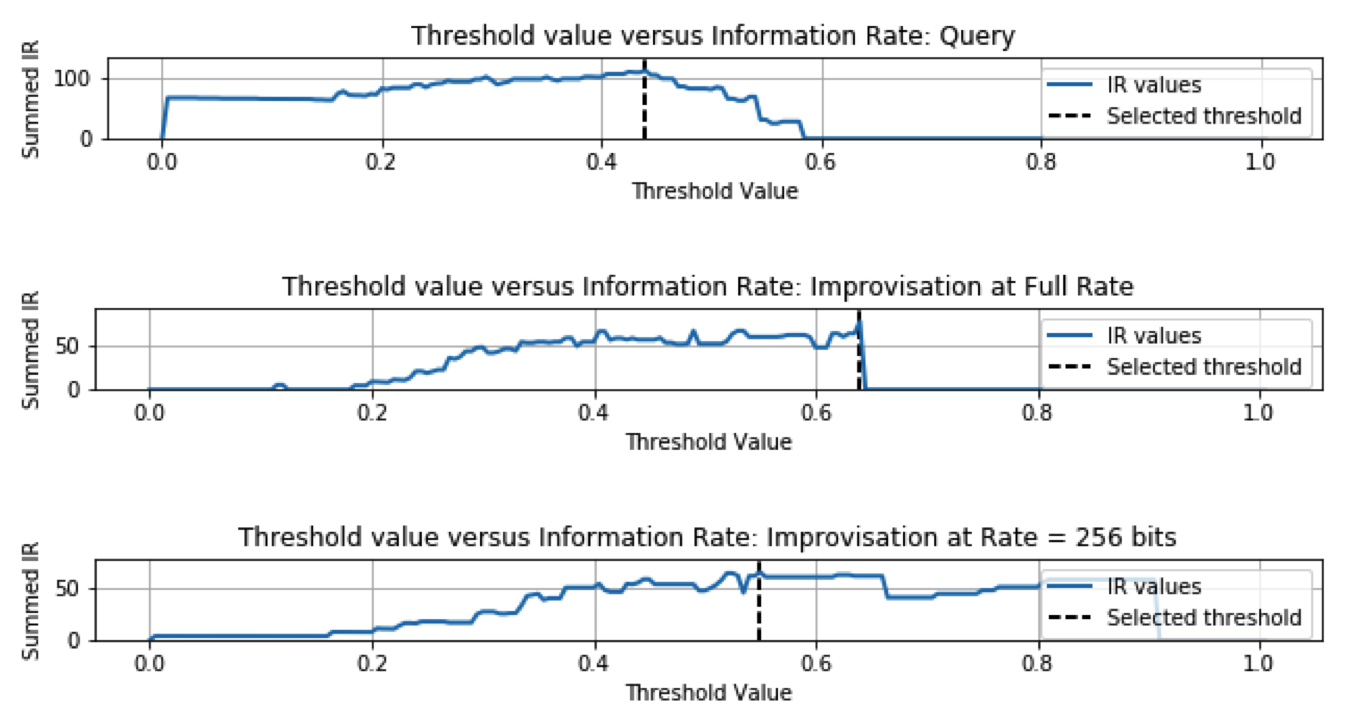}
        \caption{Information rate as function of similarity threshold found in the query (Top), full rate (Middle) and bit-rate limited resynthesis (Bottom). See text for more detail.}
        \label{fig:IR-Thresh}
    \end{figure}

    In order to asses the effect of noisy channel on VAE encoding, we performed analysis of the original query and the outputs at different bit rates. Figure \ref{fig:IR-Motifs} shows the motifs detected at optimal quantization level based on VMO analysis of the midi signal. We also plot the values of information rate as a function of similarity threshold in Figure \ref{fig:IR-Thresh}. These graphs provide an interesting insight into the level of signal detail versus repetition structure, with lower threshold corresponds to finer musical detail, usually resulting in shorter motifs. Only the motrif found at optimal threshold are shown in Figure \ref{fig:IR-Motifs}. 


    One should also note that computation of similarity requires specification of a distance function. In the results shown here we used a variant of Tonnetz distance that defines a distance on chromagram vectors (12 bin pitch-class arrays). This distance is computed by first projecting pitch vectors into a 6-dimensional space replicating the circle of pure fifth and both circles of minor and major thirds, then computing the Euclidian distance between those two (normalized) vectors \cite{tonnetz}.
    
\section{Summary and Discussion}
    In this paper we explore several theoretical relations between representation information rate connecting musical surface and latent states that are learned by a variational auto-encoder, and the predictability of sequence of these latent states over time through music information dynamics analysis of temporal information rate. We conducted several experiments where the information content of the decoder output was reduced by adding a noisy channel between the encoder and the decoder. By changing the bit-rate of the encoder-decoder we were able to move between more and less meaningful latent space representations. The synthesis-by-query process was done by encoding an input signal using a pre-trained encoder and degrading it in a controlled manner by application of bit-rate reduction before passing it to the decoder. One should note that unlike $\beta$-VAE and InfoVAE that control information rate during the phase of model training, our study focused on rate control between encoder and decoder in a pre-trained VAE. This rate control could not of course improve the representation learning itself, but allowed us to alter the extent of query influence on the output contents and its predictability in time.
    
    \subsection{Composition using VAE}
    As mentioned in the introduction, one of the motivations for investigating the information rate relations between input and output data in a pre-trained network is to use these parameters as high level criteria for composition. In order to get more musical insight into the effect of rate-control in VAE encoding-decoding, we used VMO analysis to examine the resulting musical structures. Query-based VMO resynthesis was proposed in \cite{VMO-Query}. 
    Since VMO uses human engineered musical features, it is capable of capturing more perceptually relevant musical relations, but these feature representations are not invertible or decodable directly into music. Accordingly, in previous work where VMO was used for improvisation, music generation was performed by concatenation and remixing of excerpts from the original musical data. This makes VMO impractical as a model for a huge corpus. We are looking to combine the predictive information aspect of VMO and free-energy modeling of music in a future work.

\section*{Acknowledgment}
The author would like to thank the anonymous reviewers for the thorough and informative review.

\bibliographystyle{mume}
\bibliography{my_mume2019}

\appendix
\section*{Appendix: Supplemental Materials}

\section{VAE latent information bounds}
In VAE model we denote the input as $x$ and the latent representation as $z$. The mapping of $x$ to $z$ is done by a stochastic encoder $q(z|x)$ that induces a joint distribution $p_e(x,z)=p(x)q(z|x)$. Using the definition of mutual information
\begin{eqnarray}
     I_e(x,z) & = D_{KL}(p_e(x,z)||p(x)q(z)) \\ \nonumber
     & = \mathbb{E}_{p(x)}(D_{KL}(q(z|x)||q(z)) 
\end{eqnarray}
it becomes evident that when the latent states are independent of the input, the mutual information between $x$ and $z$ is zero. In other words, for the encoding to be informative, the distribution of latent states has to substantively deviate from it marginal distribution when input signal is provided. 
Using the terminology of \cite{BrokenELBO},
\begin{equation}
H = -\int p(x) \log p(x) dx
\end{equation}
is the entropy of the musical surface,
\begin{equation}
    D = -\mathbb{E}_{p_e(x,z)} (\log p(x|z))
\end{equation}
represents the quality of reconstruction measured as negative mean log-likelihood of the decoded surface over all possible (input,latent encoding) pairs, and the rate is defined as
\begin{equation}
    R = \mathbb{E}_{p(x)}(D_{KL}(q(z|x)||p(z)),
\end{equation}
with the intuition being that this is an extra cost of representing $z$ using the encoder distribution instead of its true distribution $p(z)$ averaged over all possible inputs.

To compare this with ELBO expression, we use the definition
\begin{equation}
     ELBO : =   \mathbb{E}_{q(z|x)}(\log p(x|z)) - D_{KL}(q(z|x)||p(z)),
\end{equation}
to get an inequality $\log p(x) \geq ELBO $, where the derivation is done by writing the marginalization $p(x) = \int p(x,z) dz$ and using convexity of the log function after plugging into the integral the identity $\frac{q(z|x)}{q(z|x)}$ and finally expanding it into the expression above. 

It is implicit in this derivation that in order to estimate the distribution of our data $x$, we assume that $x$ is produced by a latent model with $z$ as a latent variable and $p(x,z)=p(x)p(z|x)$ as a joint distribution. Then we  approximate this probability by using an encoder $q(z|x)$ with $p_e(x,z)=p(x)q(z|x)$, and optimize the encoder parameters so as to find the tightest or highest lower limit to $\log p(x)$.

Averaging both sides of the ELBO inequality over all possible inputs we get
\begin{align}
   H & = -\mathbb{E}_{p(x)} \log p(x) \leq \mathbb{E}_{p(x)}(-ELBO)\\ \nonumber
   & = -\mathbb{E}_{p_e(x,z)}(\log p(x|z)) + \mathbb{E}_{p(x)} D_{KL}(q(z|x)||p(z)) \\ \nonumber
   & = D + R
\end{align}

Also noticing that  
\begin{align}
I_e(x,z) & = \mathbb{E}_{p_e(x,z)} \log \frac{q(x|z)}{p(x)} \\
& \geq \mathbb{E}_{p_e(x,z)} \log \frac{p(x|z)}{p(x)} = H - D
\end{align}

and finally
\begin{align}
    & R = \mathbb{E}_{p(x)}D_{KL}(q(z|x)||p(z)) \\ \nonumber
    & = \mathbb{E}_{p(x)} {D_{KL}(q(z|x)||q(z)) + D_{KL}(q(z)||p(z))} \\ \nonumber
    &\geq \mathbb{E}_{p(x)}(D_{KL}(q(z|x)||q(z)) = I_e(x,z),
\end{align}
we have $ H \leq I_e(x,z) + D \leq R + D = \mathbb{E}_{p(x)}(-ELBO)$, with right side equivalence existing only when the marginal distribution of the encoding equals to the "true" distribution of the latent variables that produced the surface $x$. 

One can also note that Shannon's lower bound on channel encoding $I_e(x,z)$, which is traditionally used as the lower bound on channel rate in Rate-Distortion theory, is indeed less then the rate R. 
A special consideration needs to be given to the left side of this inequality. It seems wrong that mutual information between two variables $x$ and $z$ is higher then the entropy of the individual variables. Indeed, for true distribution $p(x,z) = p(z|x)p(x) = p(x|z)p(z)$, we have $I(x,z) = H(x) - H(x|z) = H(z) - H(z|x)$, and the individual entropies are larger then or equal to mutual information due to non-negativity of the conditional entropy. In our case, the channel mutual information $I_e(x,z)$ is higher then the entropy of the data, which is in turn higher then the true mutual information between the latent states and the musical surface. 
   
\section{Deep Music Information Dynamics}
     
In order to combine the temporal and latent information rate, we can expand on using rate-distortion as follows: Instead of finding the most efficient or the lowest rate latent states up to a specified distortion, we are considering the lowest rate latent state representation that gives the best predictability into the future. We rewrite the rate-distortion objective using statistical similarity $D_{KL}(p(Y|X)||p(Y|Z))$ as our new distortion measure. This measure prefers latent states $Z$ that carry same information, or share the same belief about the future $Y$ as does the past $X$. Since $Z$ is an encoding derived from $X$, knowledge of $X$ supersedes that of $Z$ resulting in the following relations $p(Y|X,Z) = p(Y|Z)$, which establishes $Z-X-Y$ Markov chain relation between these three variables. Averaging over all possible $X,Z$ pairs we get 
\begin{align}
   \langle & D_{KL}(p(y|x)||p(y|z)) \rangle_{p(x,z)} \\
    & = \int p(x,z) p(y|x) log \frac{p(y|x)}{p(y|z)} dxdydz \\
    & = \int p(x,z,y) log \frac{p(y|x)p(x)}{p(x)p(y)}\frac{p(y)p(z)}{p(y|z)p(z)} dxdydz \\
    & = I[X,Y]-I[Z,Y] = I[X,Y|Z]
\end{align}
So what we have here is a probabilistic version of rate-distortion in time-latent space written as minimization problems of 
\begin{equation}
    \mathcal{L}=I(X,Z)+\gamma I(X,Y|Z) = I(X,Z) - \gamma I(Z,Y),
\end{equation}
where $I(X,Y)$ is neglected as it is independent of $p(Z|X)$. 
This derivation suggests a new training criteria for musical neural models that combines both latent and temporal information as 
\begin{equation}
    \mathcal{L} = I(X,Z) + \beta \langle d(X, Z) \rangle - \gamma I(Z,Y)
\end{equation}
The optimization of $\mathcal{L}$ promises finding a solution that simultaneously minimizes the rate between musical surface and latent state and maximizes the mutual information between the latent states and future of the musical surface, while balancing the distortion between the latent states and the reconstructed surface.

\subsection{Relation to VMO analysis}
As explained earlier, Information Rate of a signal quantifies the amount of information passing over time between the current measurement $Y$ and it's past $X$, formally defined in terms of the mutual information
\begin{equation}
    IR=I(Y,X) = H(Y)-H(Y|X)
\end{equation}
Maximal value of IR is obtained for signals that look instantaneously complex, thus having a high entropy $H(Y)$, but that are predictable or have little uncertainty when the signal past is taken into account, thus having low $H(Y|X)$.  In earlier studies of music information dynamics, VMO algorithm was applied to sequences of human engineered musical features. VMO is an extension of a string matching algorithm to find approximate repeated suffixes over a metric space. These suffixes are used to encode repeating blocks of the signal resulting in compression gain that is used as an estimate of the signal IR. Since the ability to find repeated suffixes depends on the level of similarity, an exhaustive search is done over all possible similarity threshold values to select the highest level of IR.

\begin{figure}
    \centering
    \includegraphics[width=\linewidth]
    {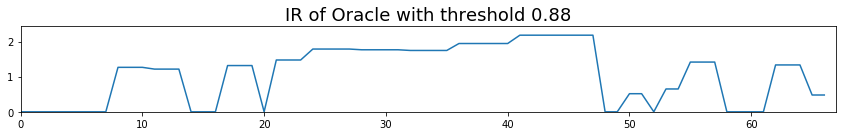}
    \caption{Information rate of the latent states at full rate (top) and at 256 bits / frame (bottom).}
    \label{fig:Z-IR}
\end{figure}

In order to gain better insight into the structure of the noise degraded VAE, we analyzed IR over the latent states sequence. In this case we choose $Y$ to be the future of the latent sequence and not future of the musical surface. We perform separate steps of training VAE using ELBO, which amounts to rate-distortion training with $\beta=1$, and then modifying the decoding rate using our noise channel. Finally the Information Rate over latent state is computed using the VMO with cosine distance between the latent vectors. 
Figure \ref{fig:Z-IR} shows IR values over time, with high IR corresponding to points on the query time-line having longer repetitions. These results illustrate that change in bit-rate indeed affects the information dynamics structure of the query, but it is hard to know exactly what musical aspects remained salient after bit-reduction.

\begin{figure}
    \centering
    \includegraphics[width=\linewidth]
    {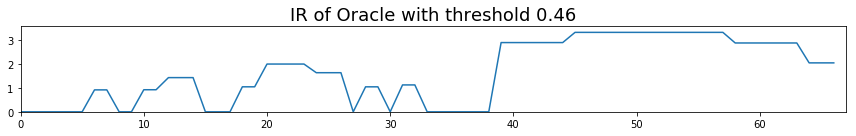}

\end{figure}

\section{Implementation details}
Musical data was represented as a piano-roll using midi-manipulation code by Dan Shiebler, available at \url{https://github.com/dshieble/Music_RNN_RBM}.

The code used to conduct these experiments and generation examples are provided at \url{https://github.com/sdubnov/qbdi}.

It uses the vmo package, available at \url{https://github.com/wangsix/vmo}.

\end{document}